# Nematic Ising superconductivity with hidden magnetism in few-layer 6$R$-TaS$_2$

Shao-Bo Liu[1†], Congkuan Tian[1,2†], Yuqiang Fang[3†], Hongtao Rong[4†], Lu Cao[7], Xinjian Wei[2], Hang Cui[1], Mantang Chen[1], Di Chen[2], Yuanjun Song[2], Jian Cui[2], Jiankun Li[2], Shuyue Guan[1], Shuang Jia[1], Chaoyu Chen[4], Wenyu He[5], Fuqiang Huang[3*], Yuhang Jiang[7*], Jinhai Mao[8], X. C. Xie[1,6,10], K.T. Law[11], Jian-Hao Chen[1,2,9,10*]

[1] International Center for Quantum Materials, School of Physics, Peking University, Beijing, China

[2] Beijing Academy of Quantum Information Sciences, Beijing, China

[3] School of Materials Science and Engineering, Shanghai Jiao Tong University, Shanghai, China

[4] Shenzhen Institute for Quantum Science and Engineering and Department of Physics, Southern University of Science and Technology, Shenzhen, China

[5] School of Physical Science and Technology, ShanghaiTech University, Shanghai, China

[6] Institute for Nanoelectronic Devices and Quantum Computing, Fudan University, Shanghai, China

[7] College of Materials Science and Optoelectronic Technology, Center of Materials Science and Optoelectronics Engineering, University of Chinese Academy of Sciences, Beijing, China

[8] CAS Center for Excellence in Topological Quantum Computation, University of Chinese Academy of Sciences, Beijing, China

[9] Key Laboratory for the Physics and Chemistry of Nanodevices, Peking University, Beijing, China

[10] Hefei National Laboratory, Hefei, China

[11] Department of Physics, Hong Kong University of Science and Technology, Hong Kong, China

[†]These authors contributed equally to this work.

[*]Corresponding Authors: Jian-Hao Chen (chenjianhao@pku.edu.cn), Fuqiang Huang (huangfq@sjtu.edu.cn), Yuhang Jiang (yuhangjiang@ucas.ac.cn)

**Abstract**

In van der Waals heterostructures (vdWHs), the manipulation of interlayer stacking/coupling allows for the construction of customizable quantum systems exhibiting exotic physics. An illustrative example is the diverse range of states of matter achieved through varying the proximity coupling between two-dimensional (2D) quantum spin liquid (QSL) and superconductors within the TaS$_2$ family. This study presents a demonstration of the intertwined physics of spontaneous rotational symmetry breaking, hidden magnetism, and Ising superconductivity in the three-fold rotationally symmetric, non-magnetic natural vdWHs 6$R$-TaS$_2$. A distinctive phase emerges in 6$R$-TaS$_2$ below a characteristic temperature ($T^*$) of approximately 30 K, which is characterized by a remarkable set of features, including a giant extrinsic anomalous Hall effect (AHE), Kondo screening, magnetic field-tunable thermal hysteresis, and nematic magneto-resistance. At lower temperatures, a coexistence of nematicity and Kondo screening with Ising superconductivity is observed, providing compelling evidence of hidden magnetism within a superconductor. This research not only sheds light on unexpected emergent physics resulting from the coupling of itinerant electrons and localized/correlated electrons in natural vdWHs but also emphasizes the potential for tailoring exotic quantum states through the manipulation of interlayer interactions.

**Introduction**

2D transition metal dichalcogenides (TMDs) have attracted considerable interest as fascinating quantum materials, exhibiting intricate interplay between charge, spin, and lattice degrees of freedom with strong spin-orbit coupling (SOC)[1,2]. This unique combination leads to the emergence of multiple intertwined orders with similar energies, giving rise to exotic physics that holds great potential for next-generation quantum devices[1-3]. In particular, the assembly of TMDs into vdWHs offers exciting prospects for realizing novel electronic ground states that are not present in the individual constituent layers[2,4-8]. Some of these states include nematicity, unconventional Ising superconductivity and chiral spin liquids[4,6,9-11].

Ising superconductivity in TMDs, such as $2H$-$MoS_2$, is generally in-plane isotropic[12]. A few exceptions, such as few-layer $1T_d$-$MoTe_2$, $2H$-$NbSe_2$ and $2M$-$WS_2$, possess in-plane anisotropic Ising-like superconductivity[13-15]. The mechanisms underlying this anisotropy have been under debate, with proposals such as anisotropic SOC[13], competing superconducting order parameters[14] and spin-orbit-parity coupling[15]. Hence, data from new in-plane anisotropic Ising superconductors would help to unravel this puzzle. Another important physical effect, the AHE, typically found in magnetic systems with spin dependent scattering[16], in topological materials with nontrivial Berry phase[17], or in systems with non-coplanar spin structures[18,19], could be an indication of time reversal symmetry breaking (TRSB). Recent reports of giant extrinsic AHE and TRSB in non-magnetic kagome metals $AV_3Sb_5$ (refs.[20-23]) and $ScV_6Sn_6$ (refs.[24,25]) suggest the involvement of chiral charge order induced loop currents. Thus, unveiling novel AHE in non-magnetic systems is of great importance to advancing scientific understanding in this field.

The non-magnetic TMDs $6R$-$TaS_2$ represents a promising natural vdWH[26,27], with alternating 2D layers of trigonal-prismatic $1H$-$TaS_2$ and octahedral $1T$-$TaS_2$. Single layer $1H$-$TaS_2$ is shown to exhibit SOC-driven Ising superconductivity[28] with a conventional s-wave order parameter[29-31], while single layer $1T$-$TaS_2$ is proven to be a correlated Mott insulator hosting a gapless QSL ground state[32,33]. Below 200 K, single layer $1T$-$TaS_2$ undergoes a $\sqrt{13} \times \sqrt{13}$ CDW transition, leading to the formation of a "start-of-David" deformation of the Ta atoms with localized electrons on an emergent frustrated triangular superlattice[32-34]. The electron spins of this superlattice are devoid of magnetic ordering due to strong frustration and quantum fluctuations, creating a QSL[5,34] (Fig. 1b). Thus, differently stacked 1$H$ and 1$T$ layers in the TaS$_2$ family become a fruitful source of exotic physics. Recent spectroscopic studies have revealed Kondo coupling between the itinerant (1$H$) and the localized (1$T$) electrons in MBE grown single layer 1$H$/1$T$ heterostructures of TaS$_2$ and TaSe$_2$ (refs.[5,8]). However, in another natural vdWHs 4$H_b$-TaS$_2$, Kondo coupling and nematic Ising superconductivity is absent deep in the superconducting state[6,35-38]. Furthermore, transport signature of intertwined nematicity and local time-reversal symmetry breaking in 1$H$/1$T$ TaS$_2$ vdWHs has been elusive so far.

In this work, we provide the first experimental evidence of the emergence of a hidden order in 6$R$-TaS$_2$ thin flakes below a characteristic temperature $T^* \sim 30$ K. Such hidden order simultaneously breaks local time reversal symmetry and in-plane rotational symmetry, and persists below the superconducting transition temperature. The coexistence of the hidden order with Ising superconductivity produces strongly nematic Ising superconductivity with simultaneous Kondo screening in the 1$T$-layer. The origin of the nematic Ising superconductivity is narrowed down to anisotropic Ising SOC[13], unconventional mixed pairing channels[14], or nematicity from the normal state; while the hidden magnetism is attributed to proximity coupling between 1$H$ and 1$T$ TaS$_2$ layers[4,39]. Our study opens new frontiers in the exploration of intertwined physics of nematicity, hidden magnetism and unconventional Ising superconductivity in non-magnetic TMDs vdWHs. A summary of the variety of interesting properties of the TaS$_2$ family with different interlayer coupling can be found in Supplementary Information S1.

Results

Nematic Ising superconductivity in thin flake 6$R$-TaS$_2$

The crystal structure of 6$R$-TaS$_2$ consists of alternating 1$H$-TaS$_2$ and 1$T$-TaS$_2$ layers, which exhibits three-fold rotational ($C_3$) symmetry along the $c$-axis with broken inversion symmetry[26] (Fig. 1a). The magneto-transport properties of superconducting and normal state 6$R$-TaS$_2$ were studied in various sample thicknesses. The device fabrication process and characterization can be found in Methods and Supplementary Information S2-S6. Three 6$R$-TaS$_2$ thin flake devices, device #1 (with thickness of 43.5 nm), device #2 (20.7 nm), device #3 (7.9 nm) as well as bulk crystals were investigated.

We found that all the thin flake devices exhibit 2D nematic Ising superconductivity. Data from device #1 is shown in Figs. 1c-e as well as in Supplementary Figs. S7-S8. Similar data from devices #2 and #3 can be found in Supplemental Figs. S8-S11. Fig. 1c plots the temperature dependent resistivity of device #1 at zero magnetic field.

Several kinks at around 300 K represents the various CDW transitions of the sample[26,27], and the superconductivity transition at around 2 K is shown in the inset. Fig. 1d is the polar plot of the normalized in-plane resistance $R_{ab}/R_N$ of device #1 from below $T_c$ to above $T_c$ with rotating in-plane magnetic field $B_{//} = 5$ T, where $\alpha$ ranges from 0° to 360° is the magnetic field angle with respect to the x-axis. Two-fold ($C_2$) nematicity in the superconducting state is evident, with minima of $R_{ab}/R_N$ appearing at $\alpha = 90°$ and 270° and maxima at $\alpha = 0°$ and 180°. Angular-dependent in-plane upper critical field $H_{c2}^{//}(\alpha)$ (Fig. S7e) and angular-dependent critical current $I_c(\alpha)$ (Fig. S7f) of device #1 also shows the same nematic behavior of $R_{ab}/R_N$. Furthermore, this anisotropy is pronounced in the superconducting state ($T < T_c^{onset} \sim 2K$), while it becomes undetectable at $B_{//} \leq 5$ T immediately above $T_c^{onset}$ (Fig. 1c). The above evidence unambiguously indicates that the two-fold anisotropy arises from nematic superconducting state rather than from resistance anisotropy[9].

Fig. 1e shows the $H_{c2}^{//}/H_p$ versus reduced temperature $T/T_c$ for $\alpha = 0°$ and 90° in 6$R$-TaS$_2$ devices #1-#3, where $H_p$ is the Pauli limit. In all the devices, $H_{c2}^{//}$ exceeds the Pauli limit for both $B_{//}$ directions, demonstrating nematic Ising superconductivity in the materials. The in-plane anisotropy obtained from $H_{c2,max}^{//}/H_{c2,min}^{//}$ ranges from ~ 2.6 to 6.5, which is much larger than previous results of Ising-like superconductors[13-15]. Additionally, nematic behaviour is observed throughout the superconducting state of 6$R$-TaS$_2$, in contrast to the isotropic behaviour observed in the ultra-low $T$ regime of another vdWHs 4$H_b$-TaS$_2$ (ref.[6]). We have ruled out trivial origins of the $C_2$ behaviors, including current induced vortex motion[40], strain stress and misalignment of magnetic field (details in Supplementary Information S13-S14), leaving the only possibility of nematic Ising superconductivity. Together with evidence shown later in the text, we argue that this state emerges with the hidden order formed at the normal state of the material.

**Intertwined physics of nematicity and hidden magnetism below $T^* \sim 30$ K**

Above the superconducting temperature and below the CDW transition temperature, a hidden order is found in 6$R$-TaS$_2$ below $T^* \sim 30$ K, which is characterized by the concurrent emergence of nematicity and hidden magnetism. Fig. 2a shows the derivative resistance (d$R_{ab}(T)$/d$T$) of device #1 at zero magnetic field, exhibiting an abrupt change at around 30 K. Fig. 2b presents the longitudinal $MR$ of device #1 under 14 T of out-of-plane magnetic field $B_\perp$, together with the two-fold anisotropic $MR$ ($AMR = \Delta R/R_{min}(\alpha)$) under 14 T of in-plane magnetic field $B_{//}$. Although $MR$ and $AMR$ are from completely different measurement configurations, we found that they could be easily scaled together and both exhibit an upturn at around 30 K. Fig. 2c plots the Hall resistivity of device #1 at various temperatures from 3 K to 80 K, with nonlinear Hall curves apparent at low temperatures. Since such nonlinear Hall curves mainly exhibits two sections with different slops (Hall resistance $R_H$), we plot $R_H$ of the low and high magnetic field regions in Fig. 2d, where a bifurcation of the two $R_H$ is found below $T^* \sim 30$ K. Fig. 2e shows the anomalous Hall conductivity $\sigma_{AHE}$ versus temperature, which again exhibits an upturn below $T^* \sim 30$ K. Fig. 2f depicts $\sigma_{AHE}$ insensitive to the angles $\theta$ between the magnetic field and the $c$-axis (up to 75° for $T = 8$ K, and > 45° for $T = 21$ K and 25.5 K), proving that the observed signal is indeed AHE[20,21]. Similar data from devices #2, #3 and #bulk can be found in Supplementary Figs. S18-S20.

This is the first time AHE in non-magnetic TMDs is observed which is not associated with a CDW transition, in contrast to the CDW-induced AHE in non-magnetic kagome metals $AV_3Sb_5$ (refs.[20,21]) and $ScV_6Sn_6$ (ref.[25]). Furthermore, below $T^* \sim 30$ K, a magnetic field-tunable thermal hysteresis loop (tunable by magnetic field up to ~ 20% and independent of the $MR$) is observed in the normal state resistance curve (Fig. 2g and Supplementary Figs. S21-S22), suggesting a first-order phase transition which is not associated with conventional structural transitions, but rather, with an electronic symmetry breaking process arising from orbital[41-43] or spin[44,45] degrees of freedom. Indeed, there has been no report of structural or conventional magnetic phase transitions in $6R$-TaS$_2$ around 30 K[26,27,46-48]. The above results unveil a hidden order with intertwined nematicity and hidden magnetism below $T^* \sim 30$ K in $6R$-TaS$_2$. The coexistence of the hidden order with Ising superconductivity produces strongly nematic Ising superconductivity with simultaneous Kondo screening in the $1T$-layer.

**Giant extrinsic AHE and Kondo resonance in non-magnetic TMDs $6R$-TaS$_2$**

AHE reflects the anomalous transverse velocity of charge carriers via extrinsic (side-jump or skew scattering) and intrinsic mechanisms[16]. Fig. 3a plots $\sigma_{AHE}$ versus $\sigma_{xx}$ of device #1, #2 and #3 together with data from other materials in the literature, spanning over the side-jump ($\sigma_{AHE} \propto \sigma_{xx}^{1.6}$), intrinsic ($\sigma_{AHE} \propto$ constant), and skew scattering ($\sigma_{AHE} \propto \sigma_{xx}$) regimes[16]. The linear dependence of $\sigma_{AHE}$ versus $\sigma_{xx}$ is not obvious with a log-log scale in Fig. 3a, and can be clearly seen in Fig. 3b with a linear scale, similar to previous results in MnGe[19], gated CsV$_3$Sb$_5$ (ref.[49]) and doped Fe (refs.[16,50]), pointing to the extrinsic skew-scattering origin[16,19,49,50]. Note that AHE is rarely found in non-magnetic materials, with limited examples such as kagome metals $AV_3Sb_5$ and $ScV_6Sn_6$ (chiral-CDW effects)[20,21,25], Weyl semimetal ZrTe$_5$ (Berry curvature effects)[17], and $6R$-TaS$_2$ in this work.

In order to elucidate the origin of AHE, we performed scanning tunneling spectroscopy (STS) on $6R$-TaS$_2$ bulk crystals with $1H$ and $1T$ termination at temperatures ranging from 0.3 K to 12 K. The STS spectra on the $1H$ termination (Figs. 3c-d) reveal a normal metal to superconductor phase transition at ~ 2.5 K with a V-shaped nodal-like pairing gap $\Delta \sim 0.53$ meV, proving 6$R$-TaS$_2$ as a strong-coupling superconductor ($2\Delta/k_BT_c \sim 4.7$), in contrast to 4$H_b$-TaS$_2$ as a weak-coupling superconductor ($2\Delta/k_BT_c \sim 3.2$)[11,35,37]. More interestingly, no clear superconducting gap is found in the 1$T$ termination (Fig. 3e-3f), while Kondo screening emerges on each of the CDW centers below ~12 K (normal state) and persists down to 0.3 K (deep inside the superconducting state). The Kondo resonance in 6$R$-TaS$_2$ is due to the screening of the localized spins on the CDW sites by the metallic 1$H$ layer underneath[5], possibly constituting a Kondo lattice like the artificial 1$H$/1$T$ heterojunctions[5], which is further supported by the spatial d$I$/d$V$ spectroscopic mapping of Kondo resonance at 4.2 K and 0.3 K (Supplementary Information S26-S27). The quantitative analysis of Kondo resonance and Kondo coupling $J_K$ can be seen in the Supplementary Information S25.

Based on all the findings presented above, we propose that spin-dependent coupling between the 1$T$ and 1$H$ layers in 6$R$-TaS$_2$ induces the unconventional hidden magnetism in 6$R$-TaS$_2$ below $T^*$; the spin state of the QSL in the 1$T$ layer may be significantly altered, i.e., the highly fluctuating spins may acquire additional broken symmetries akin to the chiral spin liquid phase[4,18,51,52]. It serves as magnetic scattering impurities without overall net magnetization, giving rise to the large extrinsic AHE in the normal states and hidden magnetism persisting down to the superconducting state in 6$R$-TaS$_2$.

**Phase diagram of nematicity, hidden magnetism and Ising SC in 6$R$-TaS$_2$**

A global phase diagram of the nematicity, hidden magnetism and unconventional Ising superconductivity in 6$R$-TaS$_2$ crystals is constructed with our experimental findings (Fig. 4a). The phase diagram can be divided into three regions: the isotropic region (gray color), the nematic normal-state region with hidden magnetism (blue color), and the nematic Ising superconductivity region with Kondo screening (red color). The first region starts from high temperature down to 30 K, where the crystal is effectively featureless after the CDW transitions at ~ 305 K, with isotropic in-plane $MR$ (anisotropy < 0.5%) down to 30 K. The second region is the normal state below $T^* \sim 30$ K, where nematicity and hidden magnetism emerge simultaneously. Here nematicity is characterized by in-plane $AMR$ as defined in Fig. 2b; hidden magnetism is characterized by 1) large extrinsic AHE under $B_\perp$, 2) magnetic field-tunable thermal hysteresis and 3) Kondo screening as shown in Figs. 3&4. The third region is the nematic Ising superconductivity coexisting with Kondo screening at the lower left corner of the phase diagram, defined as the resistivity of the sample below 90% of normal state resistivity $R_N$ (purple dots in Fig. 4a), under temperature-dependent critical in-plane fields along $\alpha = 90°$. Since the orientation of the nematicity is the same for the normal state and the superconducting state, we can conclude that nematicity of the hidden order coexist with Ising superconductivity and substantially modified the superconducting gap in different in-plane directions, producing up to ~10000% of anisotropy in resistivity. More intriguingly, when a small out-of-plane magnetic field is applied to suppress the superconductivity, the AHE effect is immediately visible without any intermediate state (Supplementary Fig. S28), together with the Kondo screening detected by STS below the superconductivity temperature, proving the coexistence of hidden magnetism and nematic Ising superconductivity in the material.

Some key points are worth noting from the phase diagram (Fig. 4a). First, the hidden phase in $6R$-TaS$_2$ ($T^* \sim 30$ K), driven by the particular coupling between the $1T$ and $1H$ layers in the $6R$ phase, is different from the spin liquid phase proposed in bulk $1T$-TaS$_2$ which persists up to 200 K[32-34]; Second, $6R$-TaS$_2$ is the first material in the TaS$_2$ family to exhibit an AHE, which interestingly echoes with theoretical prediction of AHE in triangular lattice systems with Kondo coupling of itinerant electrons and a chiral spin texture[53]; Third, both the nematic normal state and nematic Ising superconductivity reported here are firstly observed in the TaS$_2$ family, which are absent in both the single layer $1H$-TaS$_2$ (ref.[28]) and in ultra-low temperature limit of superconducting $4H_b$-TaS$_2$ (ref.[6]). Fourth, the coexistence of superconductivity and Kondo resonance in $6R$-TaS$_2$ has not been observed in either the artificial $1H/1T$ heterojunctions[5] (without SC gap on its $1H$ layer) or the natural vdWHs $4H_b$-TaS$_2$ (without Kondo resonance on its $1T$ layer) (refs.[36]). We argue that although Kondo coupling may occur in all cases (Supplementary Fig. S1), subtle differences in the coupling can lead to large variation in the spin states of the $1T$ layer as well as the electronic nematicity in the $1H$ layer, leading to marked differences in the QSL phases and other exotic physics. Thus manipulating interlayer stacking/coupling of vdWHs may help to construct customizable quantum systems with promising properties (Supplementary Fig. S1)[54]. Fourth, we note that the origin of the hidden magnetism in $4H_b$-TaS$_2$ is still under debate, including theoretical mechanisms related to CSL (ref.[52]), vison-vortex nucleation with $\mathbb{Z}_2$ topological order[55], and type-II heavy Fermi liquids[56]. For $6R$-TaS$_2$, we cannot rule out these three possible mechanisms for the origin of hidden magnetism, as the complex interactions between the $1H$ and $1T$ layers may lead to various possibilities for the spin state of $1T$. On the other hand, nematicity[57] has been observed both in the normal state of cuprates[58], iron-based[59], kagome superconductors[60], as well as in the superconducting state of magic-angle graphene[9], doped Bi$_2$Se$_3$ (ref.[61]) and NbSe$_2$ (ref.[14]). Although the microscopic origin is still unclear, nematicity may coexist, cooperate, or compete with other orders[9,57,62]. The discovery of a nematic electronic state and hidden magnetism in the normal phase $6R$-TaS$_2$ that cooperate with the Ising superconductivity points to the unconventional nature of this hidden order in the material.

**Summary**

In conclusion, a hidden order is found in 6$R$-TaS$_2$ below $T^* \sim 30$K that is characterized by simultaneous emergence of nematicity and hidden magnetism. The development of hidden magnetism may be related to significantly altered QSL state which acquire additional broken symmetries. Magneto-transport and scanning tunneling spectroscopy data strongly suggest a coexistence of hidden magnetism and nematic Ising superconductivity in 6$R$-TaS$_2$. The entangled physics of unconventional nematic Ising SC, strong-coupling SC, hidden magnetism, Kondo screening, and $\sqrt{13} \times \sqrt{13}$ CDW in 6$R$-TaS$_2$ may provide a fertile ground for the exploration of pair density waves[63,64], chiral superconductivity[11], Yu-Shiba-Rusinov (YSR)–like bound states[7], spin triplet superconductivity[65] and more. This work unveils the potential of natural van der Waals heterostructures as a promising platform for exploring the intertwined and exotic physics.

**Methods**

**Single crystals growth and characterizations**

Single crystal samples of 6$R$-TaS$_2$ have been prepared by phase transition of 1$T$-TaS$_2$ to 6$R$ at 800°C in an inert atmosphere[26,27]. The sample has been characterized by diffraction (XRD) studies in $\theta$ - $2\theta$ geometry using lab-based sources. A Quantum Design Magnetic Property Measurement System (MPMS-3) was used to measure the temperature- and field-dependent magnetization of the samples. The thickness of the various samples was measured using Atomic Force Microscopy (AFM).

**Device fabrication**

Al$_2$O$_3$-assisted exfoliation technique was used to obtain thin flakes of 6$R$-TaS$_2$ crystals with a thickness of down to 7.9 nm (more details in Supplementary Fig. S3). Standard e-beam lithography was used to pattern electrodes, followed by e-beam evaporation of Ti (5 nm) and Au (100 nm). The device fabrication process was carried out in an inert atmosphere and vacuum to minimize sample oxidation, and samples were briefly exposed to air only under PMMA capping layer protection. More details of device fabrication process can be found in Supplementary Fig. S4.

**Transport measurements**

Transport measurements were conducted at temperatures between 0.3 K and 360 K with magnetic fields up to 14 T using an Oxford Teslatron cryostat and a Quantum Design PPMS. Lock-in amplifiers were used to measure longitudinal resistance ($R_{xx}$) and Hall resistance ($R_{xy}$) at a frequency of 17.77 Hz with an AC current of 10 $\mu$A for nano-devices and 2 mA for the bulk devices. Changing the magnetic field direction was achieved by rotating the sample holder. To eliminate the influence of the slight Hall signals on the raw data of angular dependence of resistivity, the resistivity taken at every angle has been averaged with positive and negative magnetic fields. The in-plane resistivity was measured by the standard four-electrode method. Additionally, the $c$-axis resistance for the bulk sample was measured by the four-electrode method with the Corbino-shape-like configuration.

**STM/STS measurements**

Bulk 6$R$-TaS$_2$ samples were cleaved in an ultrahigh vacuum chamber at room temperature, and then immediately inserted into the STM head for further topography and spectrum measurements in continuous ultrahigh vacuum. STM measurements were performed using PtIr tips that were well calibrated on Au(111) surface. All the d$I$/d$V$ spectrum were taken using the standard lock-in technique (frequency $f$ = 973 Hz) with a small AC modulation ($V_{mod}$ = 0.05mV) added to the DC bias.

**Data availability statement**

Source data are provided with this paper. Data for figures that support the current study are available at doi://xxx.

**Acknowledgements**

This project has been supported by the National Key R&D Program of China (Grant No. 2019YFA0308402), the Innovation Program for Quantum Science and Technology (Grant No. 2021ZD0302403), the National Natural Science Foundation of China (NSFC Grant Nos. 11934001, 92265106, 11774010, 11921005, 11974347, 12074377), Beijing Municipal Natural Science Foundation (Grant No. JQ20002). Shanghai Rising-Star Program (23QA1410700). J.-H.C. acknowledges technical supports form Peking Nanofab.

**Author contributions**

J.-H.C. & S.L. conceived the idea and designed the experiments; S.L. performed all the transport measurements. S.L. & Y.S. performed the AFM measurements; C.T. performed XRD & SQUID measurements; Y.F., F.H., H.R. & C.C. provide high quality crystals; S.L. & H.C. fabricated the devices with the help of X.W. and C.T.; C.T., J.C.

and J.L. aided the transport measurements; Y.J., J.H.M. & L.C. performed the STM/STS measurements; X.C.X, W.H. & K.T.L. provided theoretical analysis; S.L. & J.-H.C. analyzed the data and wrote the manuscript; all authors commented and modified the manuscript.

**Competing interests**

The authors declare no competing interests.

**References**

1  Manzeli, S., Ovchinnikov, D., Pasquier, D., Yazyev, O. V. & Kis, A. 2D transition metal dichalcogenides. *Nat. Rev. Mater.* **2**, 17033 (2017).
2  Novoselov, K., Mishchenko, A., Carvalho, A. & Neto, A. C. 2D materials andvan der Waals heterostructures. *Science* **353**, aac9439 (2016).
3  J. M. Lu *et al.* Evidence for two-dimensional Ising superconductivity in gated MoS2. *Science* **350**, 1353-1357 (2015).
4  Persky, E. *et al.* Magnetic memory and spontaneous vortices in a van der Waals superconductor. *Nature* **607**, 692-696 (2022).
5  Vaňo, V. *et al.* Artificial heavy fermions in a van der Waals heterostructure. *Nature* **599**, 582-586 (2021).
6  Silber, I. Two-component nematic superconductivity in 4Hb-TaS2. *Nat. Commun.* **15**, 824 (2024).
7  Liu, M. K. Monolayer 1T-NbSe2 as a 2D-correlated magnetic insulator. *Sci. Adv.* **7**, eabi6339 (2021).
8  Ruan, W. *et al.* Evidence for quantum spin liquid behaviour in single-layer 1T-TaSe2 from scanning tunnelling microscopy. *Nat. Phys.* **17**, 1154-1161 (2021).
9  Cao, Y. Nematicity and Competing Orders in Superconducting Magic-Angle Graphene. *Science* **372**, 264-271 (2021).
10  Chen, M. Y., Chen, X. Y., Yang, H., Du, Z. Y. & Wen, H.-H. Superconductivity with twofold symmetry in Bi2Te3/FeTe0.55Se0.45 heterostructures. *Sci. Adv.* **4**, eaat1084 (2018).
11  A. Ribak *et al.* Chiral superconductivity in the alternate stacking compound 4Hb-TaS2. *Sci. Adv.* **6**, eaax9480 (2020).
12  Wan, P. *et al.* Orbital Fulde–Ferrell–Larkin–Ovchinnikov state in an Ising superconductor. *Nature* **619**, 46-51 (2023).
13  Cui, J. *et al.* Transport evidence of asymmetric spin–orbit coupling in few-layer superconducting 1Td-MoTe2. *Nat. Commun.* **10**, 2044 (2019).
14  Hamill, A. *et al.* Two-fold symmetric superconductivity in few-layer NbSe2. *Nat. Phys.* **17**, 949-954 (2021).
15  Zhang, E. *et al.* Spin–orbit–parity coupled superconductivity in atomically thin 2M-WS2. *Nat. Phys.* **19**, 106-113 (2022).
16  Nagaosa, N., Sinova, J., Onoda, S., MacDonald, A. H. & Ong, N. P. Anomalous Hall effect. *Rev. Mod. Phys.* **82**, 1539-1592 (2010).
17  Liang, T. *et al.* Anomalous Hall effect in ZrTe5. *Nat. Phys.* **14**, 451-455 (2018).

18    Machida, Y., Nakatsuji, S., Onoda, S., Tayama, T. & Sakakibara, T. Time-reversal symmetry breaking and spontaneous Hall effect without magnetic dipole order. *Nature* **463**, 210-213 (2010).

19    Fujishiro, Y. *et al.* Giant anomalous Hall effect from spin-chirality scattering in a chiral magnet. *Nat. Com.* **12**, 317 (2021).

20    Yang, S.-Y. Giant, unconventional anomalous Hall effect in the metallic frustrated magnet candidate, KV3Sb5. *Sci. Adv.* **6**, eabb6003 (2020).

21    Yu, F. H. *et al.* Concurrence of anomalous Hall effect and charge density wave in a superconducting topological kagome metal. *Phys. Rev. B* **104**, L041103 (2021).

22    Mielke, C. *et al.* Time-reversal symmetry-breaking charge order in a kagome superconductor. *Nature* **602**, 245-250 (2022).

23    Guo, C. *et al.* Switchable chiral transport in charge-ordered kagome metal CsV3Sb5. *Nature* **611**, 461-466 (2022).

24    Guguchia, Z. Hidden magnetism uncovered in a charge ordered bilayer kagome material ScV6Sn6. *Nat. Commun.* **14**, 7796 (2023).

25    Yi, C. Quantum oscillations revealing topological band in kagome metal ScV6Sn6. *Phys. Rev. B* **109**, 035124 (2024).

26    Achari, A. *et al.* Alternating Superconducting and Charge Density Wave Monolayers within Bulk 6R-TaS2. *Nano Lett.* **22**, 6268-6275 (2022).

27    Pal, S. Charge density wave and superconductivity in 6R-TaS2. *Physica B* **669**, 415266 (2023).

28    de la Barrera, S. C. *et al.* Tuning Ising superconductivity with layer and spin–orbit coupling in two-dimensional transition-metal dichalcogenides. *Nat. Commun.* **9**, 1427 (2018).

29    Kvashnin, Y. *et al.* Coexistence of Superconductivity and Charge Density Waves in Tantalum Disulfide: Experiment and Theory. *Phys. Rev. Lett.* **125**, 186401 (2020).

30    Li, L. *et al.* Superconducting order from disorder in 2H-TaSe2-xSx. *npj Quantum Mater.* **2**, 1-7 (2017).

31    Wan, Z. Signatures of Chiral Superconductivity in Chiral Molecule Intercalated Tantalum Disulfide. Preprint at https://arxiv.org/abs/2302.05078 (2023).

32    Law, K. T. & Lee, P. A. 1T-TaS2 as a quantum spin liquid. *Proc. Natl Acad. Sci. USA* **114**, 6996-7000 (2017).

33    He, W.-Y., Xu, X. Y., Chen, G., Law, K. T. & Lee, P. A. Spinon Fermi Surface in a Cluster Mott Insulator Model on a Triangular Lattice and Possible Application to 1T−TaS2. *Phys. Rev. Lett.* **121**, 046401 (2018).

34    Klanjšek, M. *et al.* A high-temperature quantum spin liquid with polaron spins. *Nat. Phys.* **13**, 1130-1134 (2017).

35    Nayak, A. K. *et al.* Evidence of topological boundary modes with topological nodal-point superconductivity. *Nat. Phys.* **17**, 1413-1419 (2021).

36    Wen, C. *et al.* Roles of the Narrow Electronic Band near the Fermi Level in 1T-TaS2-Related Layered Materials. *Phys. Rev. Lett.* **126**, 256402 (2021).

37    Shen, S. *et al.* Coexistence of Quasi-two-dimensional Superconductivity and Tunable Kondo Lattice in a van der Waals Superconductor. *Chin. Phys. Lett.* **39** (2022).

38    Nayak, A. K. First-order quantum phase transition in the hybrid metal–Mottinsulator transition metal dichalcogenide 4Hb-TaS2. *Proc. Natl Acad. Sci. USA* **120**, e2304274120 (2023).

39    Song, X.-Y., Vishwanath, A. & Zhang, Y.-H. Doping the chiral spin liquid: Topological superconductor or chiral metal. *Phys. Rev. B* **103**, 165138 (2021).

40    Wang, Y.-L. *et al.* Parallel magnetic field suppresses dissipation in superconducting nanostrips. *Proc. Natl Acad. Sci. USA* **114**, E10274–E10280 (2017).

41    Liang, Y. G. *et al.* Tuning the hysteresis of a metal-insulator transition via lattice compatibility. *Nat. Commun.*

**11**, 3539 (2020).

42  Haverkort, M. W. *et al.* Orbital-assisted metal-insulator transition in VO2. *Phys. Rev. Lett.* **95**, 196404 (2005).

43  Wei, Z. *et al.* Two superconductor-insulator phase transitions in the spinel oxide Li1±xTi2O4−δ induced by ionic liquid gating. *Phys. Rev. B* **103**, L140501 (2021).

44  Liuwan Zhang, Casey Israel, Amlan Biswas, R. L. Greene & Lozanne, A. d. Direct Observation of Percolation in a Manganite Thin Film. *Science* **298**, 805 (2002).

45  Catalano, S. *et al.* Tailoring the electronic transitions of NdNiO3 films through (111)pc oriented interfaces. *APL Materials* **3**, 062506 (2015).

46  Thompson, A. H. The synthesis and properties of 6R-TaS2. *Solid State Commun.* **17**, 1115-1117 (1975).

47  Figueroa, E. Physical properties of 6R-TaS2. *J. Solid State Chem.* **114**, 486-490 (1992).

48  Liu, Y. *et al.* Tuning the charge density wave and superconductivity in 6R-TaS2−xSex. *J. Appl. Phys.* **117** (2015).

49  Zheng, G. *et al.* Electrically controlled superconductor-to-failed insulator transition and giant anomalous Hall effect in kagome metal CsV3Sb5 nanoflakes. *Nat. Commun.* **14**, 678 (2023).

50  Shiomi, Y., Onose, Y. & Tokura, Y. Extrinsic anomalous Hall effect in charge and heat transport in pure iron, Fe0.997Si0.003, and Fe0.97Co0.03. *Phys. Rev. B* **79**, 100404(R) (2009).

51  Wen, X. G., Wilczek, F. & Zee, A. Chiral spin states and superconductivity. *Phys. Rev. B* **39**, 11413-11423 (1989).

52  Lin, S.-Z. Kondo enabled transmutation between spinons and superconducting vortices origin of magnetic memory in 4Hb-TaS2. Preprint at https://arxiv.org/pdf/2210.06550 (2022).

53  Martin, I. & Batista, C. D. Itinerant Electron-Driven Chiral Magnetic Ordering and Spontaneous Quantum Hall Effect in Triangular Lattice Models. *Physical Review Letters* **101** (2008).

54  Andrei, E. Y. *et al.* The marvels of moiré materials. *Nat. Rev. Mater.* **6**, 201-206 (2021).

55  Chen, G. Is spontaneous vortex generation in superconducting 4Hb-TaS2 from vison-vortex nucleation with Z2 topological order? , Preprint at https://doi.org/10.48550/arXiv.42208.03995 (2022).

56  König, E. J. Type-II heavy Fermi liquids and the magnetic memory of 4Hb-TaS2. *Phys. Rev. Res.* **6**, L012058 (2024).

57  Fradkin, E., Kivelson, S. A., Lawler, M. J., Eisenstein, J. P. & Mackenzie, A. P. Nematic fermi fluids in condensed matter physics. *Annu. Rev. Condens. Matter Phys.* **1**, 153-178 (2010).

58  Tranquada, J. M., Sternlieb, B. J., Axe, J. D., Nakamura, Y. & Uchida, S. Evidence for stripe correlations of spins and holes in copper oxide. *Nature* **375**, 561–563 (1995).

59  Chu, J.-H. In-plane resistivity anisotropy in an underdoped iron arsenide. *Science* **329**, 824–826 (2010).

60  Xiang, Y. *et al.* Twofold symmetry of c-axis resistivity in topological kagome superconductor CsV3Sb5 with in-plane rotating magnetic field. *Nat. Commun.* **12**, 6727 (2021).

61  Matano, K., Kriener, M., Segawa, K., Ando, Y. & Zheng, G.-q. Spin-rotation symmetry breaking in the superconducting state of CuxBi2Se3. *Nat. Phys.* **12**, 852-854 (2016).

62  Fernandes, R. M., Chubukov, A. V. & Schmalian, J. What drives nematic order in iron-based superconductors? *Nat. Phys.* **10**, 97-104 (2014).

63  Chen, H. *et al.* Roton pair density wave in a strong-coupling kagome superconductor. *Nature* **599**, 222-228 (2021).

64  Zhao, H. *et al.* Smectic pair-density-wave order in EuRbFe4As4. *Nature* **618**, 940-945 (2023).

65  Gu, Q. *et al.* Detection of a pair density wave state in UTe2. *Nature* **618**, 921-927 (2023).

**Figure and Captions**

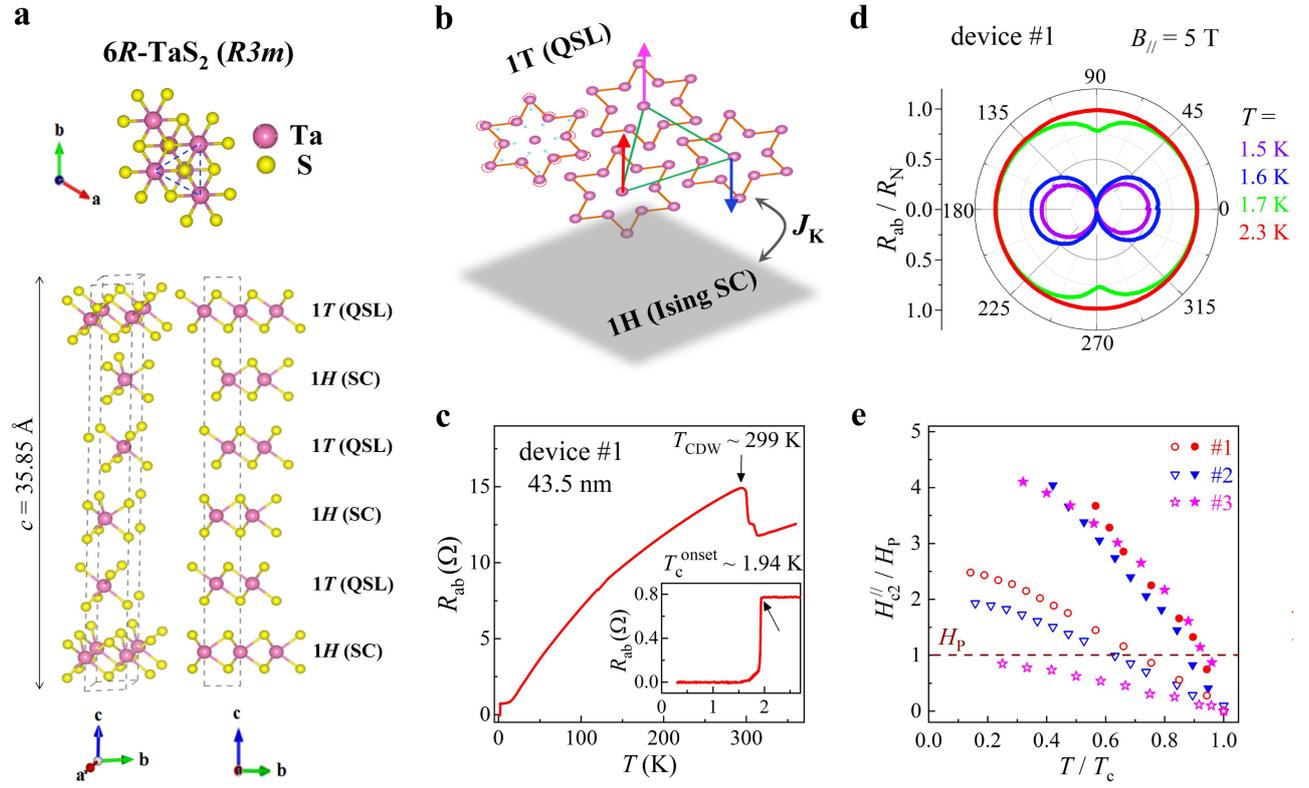

**Fig. 1. Natural vdWHs, QSL candidate and nematic Ising SC in few-layer 6$R$-TaS$_2$. a**, Crystal structure of 6$R$-TaS$_2$ in three different angle views, exhibiting three-fold rotational ($C_3$) symmetry along the $c$-axis with broken inversion symmetry. **b**, Schematics of a 1$T$/1$H$ bilayer comprising one-third of the 6$R$-TaS$_2$ unit cell. The localized moments in the 1$T$ layer interacts with the Ising superconducting 1$H$ layer through Kondo coupling ($J_K$) or proximity effect. **c**, Resistivity vs. $T$ at zero magnetic field for 6$R$-TaS$_2$ device #1 for $T$ from 0.3K to 360K. Inset shows the zoom-in view of the superconducting transition. **d**, Polar plot of angular-dependent normalized in-plane resistance $R_{ab}/R_N$ for device #1 under various $T$ near the $T_c^{onset}$ at $B_{//}$ = 5 T, where the magnetic field angle $\alpha$ with respect to $x$-axis ranges from 0° to 360°. **e**, Normalized in-plane upper critical field $H_{c2}^{//}/H_p$ vs. reduced temperature $T/T_c$ along the in-plane directions of $\alpha$ = 90° (solid dots) and 0° (empty dots). The brown dashed line denotes the normalized Pauli limit $H_p$.

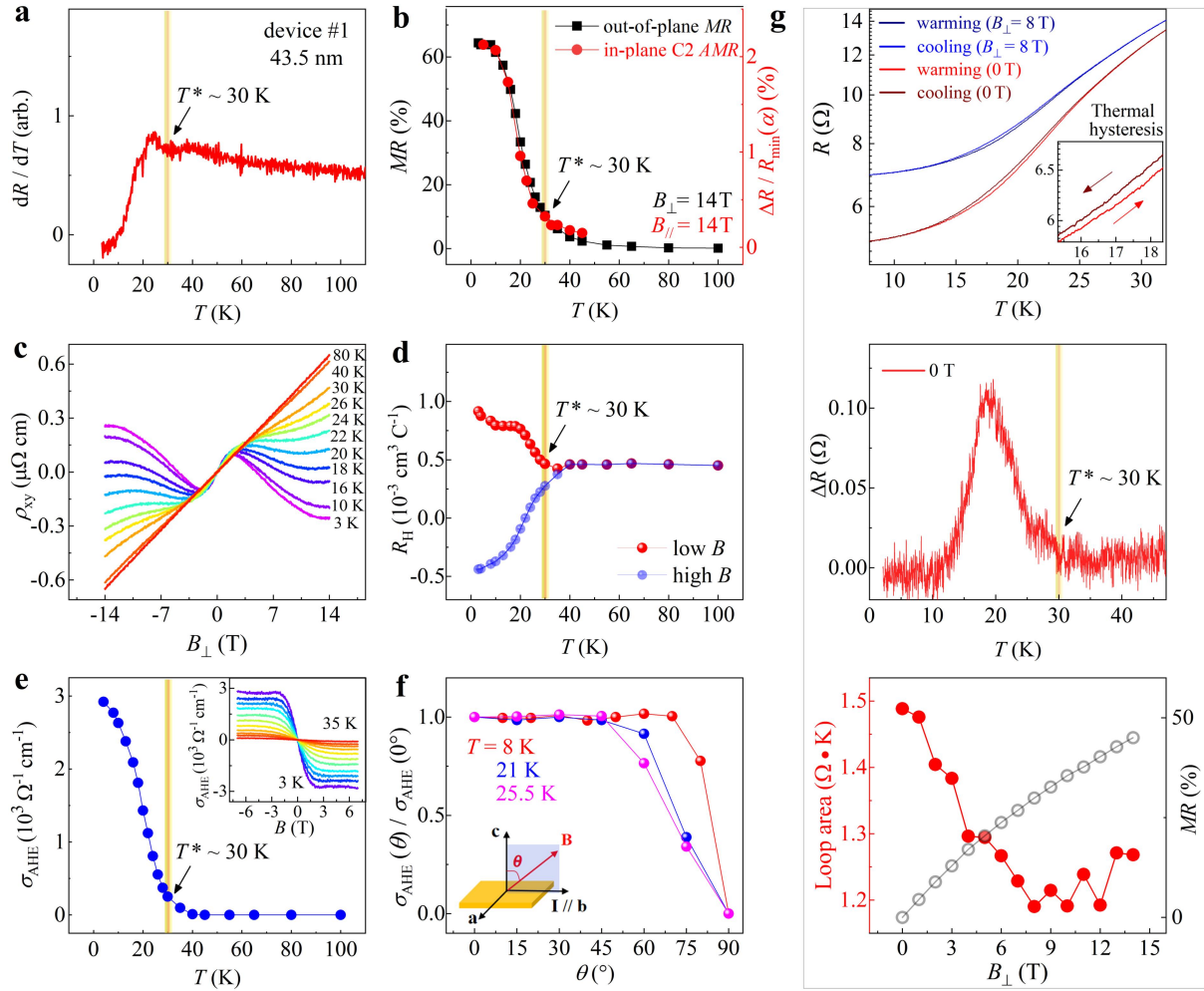

**Fig. 2. Hidden order with intertwined physics of nematicity, AHE, and magnetic field-tunable thermal hysteresis around $T^* \sim 30$ K for device #1. a**, $T$-dependent derivative resistance curve ($dR_{ab}/dT$) at zero magnetic field, showing an abrupt change around $T^* \sim 30$ K. **b**, $T$-dependent longitudinal $MR$ under $B_\perp = 14$ T (extracted from Supplementary S15) and anisotropic magneto-resistance ($AMR$) under $B_{//} = 14$ T (extracted from Supplementary S16), exhibiting a concurrent upturn at $T^* \sim 30$ K. **c**, Hall resistivity $\rho_{xy}(B_\perp)$ at various $T$ from 3 K to 80 K, with nonlinear Hall curves apparent below $T^* \sim 30$ K. **d**, $T$-dependent Hall coefficient $R_H \equiv d\rho_{xy}/dB$ extracted from both low and high magnetic field regions, exhibiting a bifurcation below $T^* \sim 30$ K. **e**, $T$-dependent anomalous Hall conductivity $\sigma_{AHE}$ (main panel) and magnetic field-dependent $\sigma_{AHE}$ (inset) at various $T$, exhibiting an upturn in $\sigma_{AHE}$ below $T^* \sim 30$ K. The Hall conductivity $\sigma_{xy} = -\rho_{xy}/(\rho_{xx}^2 + \rho_{xy}^2)$, then $\sigma_{AHE}$ obtained by subtracting the local linear ordinary Hall conductivity background. **f**, $\sigma_{AHE}(\theta)/\sigma_{AHE}(0°)$ vs. $\theta$, where $\theta$ is the angle between the magnetic field and the $c$ axis under various $T$ (extracted from Supplementary S17). Inset shows measurement configuration. **g**, Magnetic field-tunable thermal hysteresis. Upper panel: $T$-dependent resistance with decreasing and increasing $T$ under $B_\perp = 0$ T and 8 T. Inset is a zoom-in plot of the thermal hysteresis at zero field. Middle panel: $T$-dependent $\Delta R$ (difference in resistance between warming and cooling ramps), exhibiting the temperature range (between ~11 K to 30 K) of the thermal hysteresis. Lower panel: $B_\perp$-dependent integral area of the hysteresis loop and $B_\perp$-dependent longitudinal $MR$ at $T = 15$ K (extracted from Supplementary S21). The loop area is clearly independent of the $MR$.

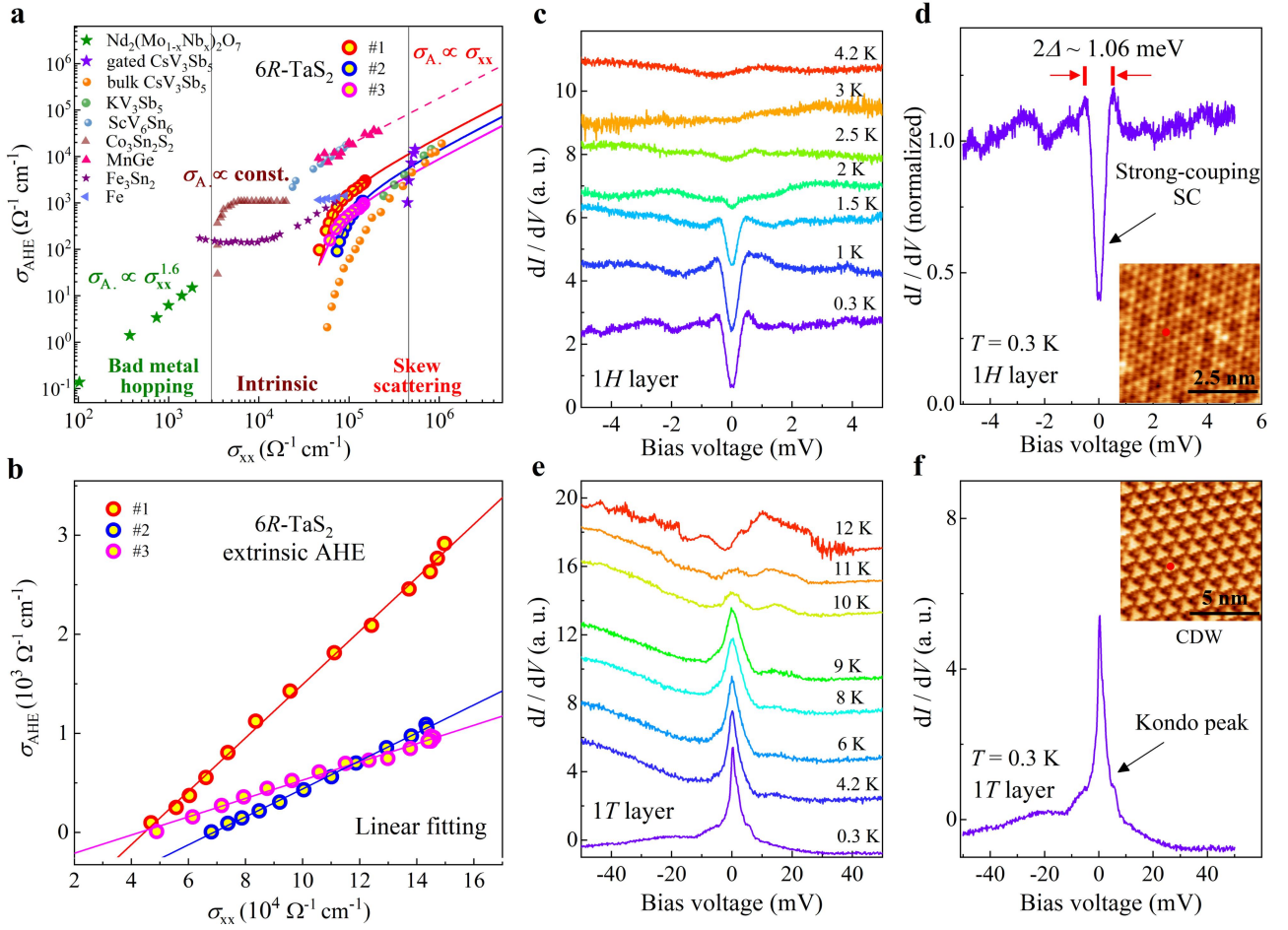

**Fig. 3. Extrinsic AHE with linear scaling relation and Kondo resonance in non-magnetic TMDs 6R-TaS₂. a**, Map of AHE ($\sigma_{AHE}$ vs. $\sigma_{xx}$) for various materials in double logarithmic coordinates, spanning over the side-jump ($\sigma_{AHE} \propto \sigma_{xx}^{1.6}$), intrinsic ($\sigma_{AHE} \propto$ constant), and skew scattering ($\sigma_{AHE} \propto \sigma_{xx}$) regimes. The solid lines are the linear fitting of $\sigma_{AHE}$ vs. $\sigma_{xx}$ for 6R-TaS₂ device #1, #2 and #3. **b**, The linear dependence of $\sigma_{AHE}$ vs. $\sigma_{xx}$ for 6R-TaS₂ samples with a linear scale. **c**, T-dependent dI/dV spectrum of the 1H layer at zero magnetic field ($V_s = -5$ mV, $I = 400$ pA). **d**, Zero-field STS spectra of the superconducting gap at $T = 0.3$ K and atomically resolved topography of the 1H layer (inset). The red dot shows the location where the STS spectra was measured. **e**, Kondo resonance in the 1T layer from 0.3 K (superconducting state) to up to 12 K (normal state) ($V_s = -50$ mV, $I = 400$ pA). **f**, Kondo resonance of the 1T layer at $T = 0.3$ K. Inset shows STM topography of the 1T layer with the $\sqrt{13} \times \sqrt{13}$ CDW pattern. The red dot shows the location where the STS spectra was measured.

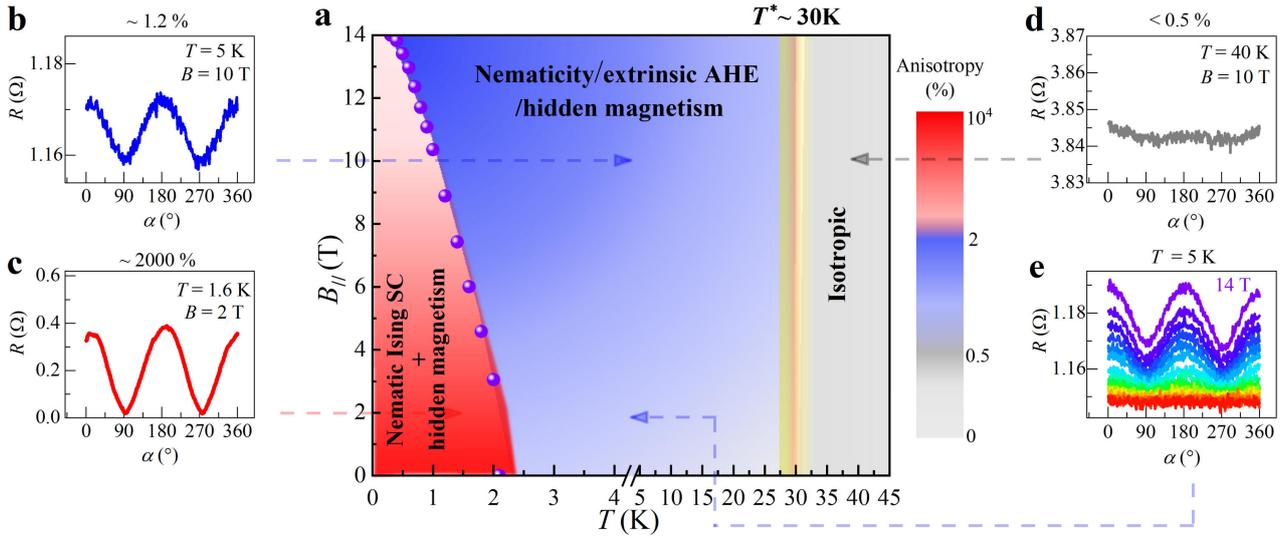

**Fig. 4. Phase diagram of nematicity, hidden magnetism and Ising superconductivity in 6$R$-TaS$_2$. a**, $B_{//}$ - $T$ phase diagram of 6$R$-TaS$_2$ (device #1), divided into three regions: the isotropic region (gray color), the weakly nematic with hidden magnetism region (blue color), and the nematic Ising superconductivity with hidden magnetism region (red color). The purple dots are the boundary between nematic Ising superconducting state and weakly nematic normal state, representing the upper critical field along $\alpha = 90°$ determined by the 90% $R_N$ criterion (extracted from Fig. S6g). The yellow ribbon around $T^*$ is the boundary between the weakly nematic and isotropic normal state regions. The color scale represents nematicity defined as $[R_{max}(\alpha) - R_{min}(\alpha)] / R_{min}(\alpha)$(%) and extracted from the *AMR* (See Supplementary S16). **b-d**, The representative in-plane *AMR* curves $R(\alpha)$ at 5 K and 10 T (blue color region), 1.6 K and 2 T (red color region), 40 K and 10 T (gray color region), respectively. **e**, In-plane *AMR* curves at 5 K measured with $B_{//}$ from 1 T to 14 T.